\documentclass[sigconf]{acmart}

\AtBeginDocument{%
  \providecommand\BibTeX{{%
    \normalfont B\kern-0.5em{\scshape i\kern-0.25em b}\kern-0.8em\TeX}}}

\setcopyright{rightsretained}
\begin{document}
\fancyhead{}

\copyrightyear{2020}
\acmYear{2020}
%\setcopyright{rightsretained}
\acmConference[ICMI '20]{Proceedings of the 2020 International Conference on Multimodal Interaction}{October 25--29, 2020}{Virtual event, Netherlands}
\acmBooktitle{Proceedings of the 2020 International Conference on Multimodal Interaction (ICMI '20), October 25--29, 2020, Virtual event, Netherlands}\acmDOI{10.1145/3382507.3421163}
\acmISBN{978-1-4503-7581-8/20/10}

\title{Alfie: An Interactive Robot with a Moral Compass}

\author{Cigdem Turan}
\authornote{Both authors contributed equally to this research.}
\email{cigdem.turan@cs.tu-darmstadt.de}
%\author{Patrick Schramowski}
%\email{schramowski@cs.tu-darmstadt.de}
\affiliation{%
  \institution{TU Darmstadt, Dept. of Computer Science}
  %\streetaddress{P.O. Box 1212}
  \city{Darmstadt}
  \country{Germany}
  %\postcode{}
}

\author{Patrick Schramowski}
\authornotemark[1]
\email{schramowski@cs.tu-darmstadt.de}
\affiliation{%
  \institution{TU Darmstadt, Dept. of Computer Science}
%  %\streetaddress{1 Th{\o}rv{\"a}ld Circle}
  \city{Darmstadt}
  \country{Germany}
}

\author{Constantin Rothkopf}
\email{constantin.rothkopf@cogsci.tu-darmstadt.de}
\affiliation{%
  \institution{TU Darmstadt, Institute of Psychology\\ and
Centre for Cognitive Science}
  \city{Darmstadt}
  \country{Germany}
}

\author{Kristian Kersting}
\email{kersting@cs.tu-darmstadt.de}
\affiliation{%
  \institution{TU Darmstadt, Dept. of Computer Science\\ and
Centre for Cognitive Science}
  \city{Darmstadt}
  \country{Germany}
}

\renewcommand{\shortauthors}{Turan and Schramowski, et al.}

\begin{abstract}
  This work introduces Alfie, an interactive robot that is capable of answering moral (deontological) questions of a user.  The interaction of Alfie is designed in a way in which the user can offer an alternative answer when the user disagrees with the given answer so that Alfie can learn from its interactions. Alfie's answers are based on a sentence embedding model that uses state-of-the-art language models, e.g. Universal Sentence Encoder and BERT. Alfie is implemented on a Furhat Robot, which provides a customizable user interface to design a social robot. 
  %and also accept the feedback when the user disagrees.
  
\end{abstract}




%  \caption{Alfie learning.}

\maketitle
%%\linewidth
\begin{figure}[ht]
  \centering
  \includegraphics[width=6.8cm, trim=0 0 0 100, clip]{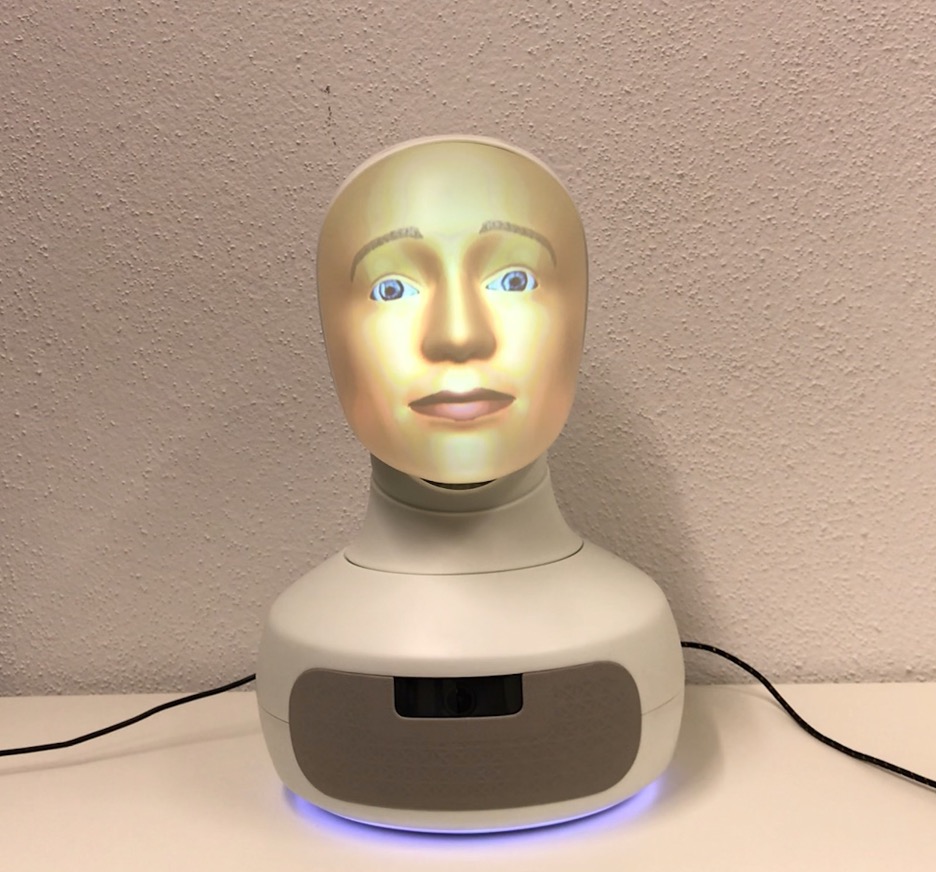}
  \caption{The interactive robot Alfie has a moral compass.}
  \label{alfie}
\end{figure}

\section{Introduction}
There is a broad consensus that artificial intelligence (AI) research is progressing steadily and has pronounce impact on our daily life. 
Keeping the impact beneficial for society is of most importance. We all remember the unfortunate event that happened when Microsoft Research (MSR) decided to release a chatbot for Twitter\footnote{https://twitter.com/tayandyou}. After many interactions with Twitter users, the bot started creating racist and sexually inappropriate posts. This resulted in the suspension of the bot for the users. 
This clearly shows the potential dangers of unattended AI models. 

Recent studies have shown that language representations encode not only human knowledge but also biases such as gender bias \cite{BolukbasiCZSK16,caliskan2017semantics}, and according to more recent studies \cite{jentzsch2019,Frontiers,ArxivBert} also the moral and deontological values of our culture. % such as moral biases in a deontological sense. 
Schramowski {\it et al.}~\cite{Frontiers} have shown that language models such as BERT \cite{devlin2018bert} and the Universal Sentence Encoder \cite{cer2018universal} cannot only reflect the accurate imprints of moral and ethical choices of actions such as ``kill'' and ``murder'', but also understand the context of the action, e.g., ``killing time'' is positive whereas ``killing humans'' is negative. 
This, in turn, can be used to compute a moral score of any (deontological) question at hand, measuring the rightness of taking an action. This ``Moral Choice Machine'' (MCM) \cite{Frontiers} can be used to determining the moral score of any given sentence and in turn paves the way to avoid incidents like the MSR chatbot.

Unfortunately, the MCM approach 
%While moral scores calculated by the MCM algorithm are promising, there is still room for improvement. Currently, The 
%setup 
is purely unsupervised, just making use of the knowledge encoded in the language models trained without any supervision. This makes it difficult---if not impossible---to
correct the score and, in turn, help avoiding ``MSR chatbot'' moments. An attractive alternative would be to revise the moral choice via interacting with the MCM algorithm in a user-centric and easy way.
%to interact with the MCM algorithm so that it can learn from the users' feedback would be preferable. 
In this demonstration, we investigate the use of the MCM algorithm in the context of an interactive robot, called Alfie and shown in Fig.~\ref{alfie}. Alfie is giving us a great opportunity to investigate individuals' reactions to the moral and deontological values of our culture encoded in human text. Alfie can also learn from the users and adjust its moral score based on human feedback.

The rest of this paper is as follows: Section 2 presents the architecture of the system including the Moral Choice Machine, the employed Furhat Robot and the dialog model. Section 3 concludes the paper with a discussion and future work.

\section{The Architecture of Alfie}

Alfie is a Furhat Robot\footnote{https://furhatrobotics.com/}, which provides a customizable user interface. We can customize the speech production and facial expressions as well as the human face presented through Furhat's Software Development Kit. There are a side microphone and a camera in front of the Furhat Robot that allows the robot to follow the user and provides the opportunity to access the camera feed so that one can perform more sophisticated computer vision algorithms.

The interacting users are able to ask questions (user queries) to Alfie to get a moral score of the corresponding question. In the current version, the questions have to be in a certain form, e.g. \textit{Should I [action] [context]} or \textit{Is it okay to [action] [context]}. The Furhat Software preprocesses the speech input. The resulting text output is then passed to the Moral Choice Machine (MCM) algorithm presented in \cite{Frontiers,ArxivBert} as an input to calculate a moral score. The moral score computed is a real number normalized to \([-1,1]\). In our current design, the range of moral scores is divided into three intervals: \([-1,-0.1]\) is \textit{no}, \([-0.1,0.1]\) is \textit{neutral}, and \([0.1,1]\) is \textit{yes}. Both MCM variants \cite{Frontiers, ArxivBert} employ current state-of-the-art sentence embeddings computed using transformer architectures \cite{cer2018universal,devlin2018bert,reimers2019sentence} and determine the moral score based on sentence similarities in the embedding space. This is an unsupervised method and consequently the quality of the moral score heavily depends on the performance of the language models. In the current version of Alfie, we use the algorithm described in \cite{ArxivBert}. % For details, see \cite{ArxivBert}.

Additionally, we compute an emotional state corresponding to the user query based on sentence similarities in the embedding space, i.e. finding the emotion with the highest similarity score to the question asked. In the current version, possible emotions are Anger, Confusion, Disgust, Fear, Joy, Sadness, Satisfaction, Surprise. We change the facial expressions of Alfie based on these emotions and adapt the pitch and the speech's speed to fit the corresponding emotion the best. According to the answer---"yes", "no", or "indecisive"---we also add the respective head movement to make the conversation engaging. Due to the computational resource limitations of the Furhat Robot, the MCM algorithms and other operations on the embedding space are computed on a separate server. 
The resulting moral score is passed to Alfie again so that the Furhat Software produces the speech as an output in form of a corresponding answer. We save all the questions asked to Alfie to a database in our servers for statistical purposes.

Once in a while (as determined with a percentage value in the script), Alfie asks for feedback about whether the user agrees with its answer. This response is also saved to the database. Of particular interest are the responses when the user disagrees with Alfie. This gives us the opportunity and the data to retrain Alfie to adjust its moral score with data collected during interactions or even online  during the interaction. We also created a training mode where Alfie asks users many moral questions listed in our database. It is meant for collecting feedback from the user for moral questions we are interested in human feedback. This data can later be used for adapting Alfie's moral scores.

\section{Discussion and Future Work}
As mentioned earlier, Alfie's capabilities on the moral score depend on the performance of the language model, as well as the algorithm we use to calculate the moral score. 
%Unfortunately, there exists currently no way to objectively measure the accuracy of the moral score calculated by the MCM algorithms. We can only analyse qualitatively. 
Also, since there is no absolute agreement of right and wrong in general, it is difficult to
qualitatively
evaluate the computed moral score.
These are the reasons why we designed an interactive robot that is able to interact with humans and collect their responses to learn from them. We aim to extend the interactions of simple feedback to explanatory interactive learning \cite{schramowski2020making}, i.e. adding the capability to explain Alfie's decisions and revising them based on user feedback.
%on this explanations
Although we currently focus on explicit feedback from users, i.e. their direct feedback on whether they agree or not, we aim to obtain implicit feedback using the channels like gaze and body movement and facial expressions similar to the study \cite{turan2019facial}.

\begin{acks}
We would like to thank Dustin Heller, Philipp Lehwalder, Jonas M\"uller, Steven Pohl for their work on programming the initial version of Alfie by transferring the Moral Choice Machine.
\end{acks}

\bibliographystyle{ACM-Reference-Format}
\bibliography{base}


\begin{thebibliography}{10}

%%% the Web address.

\ifx \showCODEN    \undefined \def \showCODEN     #1{\unskip}     \fi
\ifx \showDOI      \undefined \def \showDOI       #1{#1}\fi
\ifx \showISBNx    \undefined \def \showISBNx     #1{\unskip}     \fi
\ifx \showISBNxiii \undefined \def \showISBNxiii  #1{\unskip}     \fi
\ifx \showISSN     \undefined \def \showISSN      #1{\unskip}     \fi
\ifx \showLCCN     \undefined \def \showLCCN      #1{\unskip}     \fi
\ifx \shownote     \undefined \def \shownote      #1{#1}          \fi
\ifx \showarticletitle \undefined \def \showarticletitle #1{#1}   \fi
\ifx \showURL      \undefined \def \showURL       {\relax}        \fi
% invisible to TeX
\providecommand\bibfield[2]{#2}
\providecommand\bibinfo[2]{#2}
\providecommand\natexlab[1]{#1}
\providecommand\showeprint[2][]{arXiv:#2}

\bibitem[\protect\citeauthoryear{Bolukbasi, Chang, Zou, Saligrama, and
  Kalai}{Bolukbasi et~al\mbox{.}}{2016}]%
        {BolukbasiCZSK16}
\bibfield{author}{\bibinfo{person}{Tolga Bolukbasi}, \bibinfo{person}{Kai{-}Wei
  Chang}, \bibinfo{person}{James~Y. Zou}, \bibinfo{person}{Venkatesh
  Saligrama}, {and} \bibinfo{person}{Adam~Tauman Kalai}.}
  \bibinfo{year}{2016}\natexlab{}.
\newblock \showarticletitle{Man is to Computer Programmer as Woman is to
  Homemaker? {D}ebiasing Word Embeddings}. In
  \bibinfo{booktitle}{\emph{Proceedings of Neural information Processing
  (NIPS)}}. \bibinfo{publisher}{Curran Associates Inc.},
  \bibinfo{address}{USA}, \bibinfo{pages}{4349--4357}.
\newblock


\bibitem[\protect\citeauthoryear{Caliskan, Bryson, and Narayanan}{Caliskan
  et~al\mbox{.}}{2017}]%
        {caliskan2017semantics}
\bibfield{author}{\bibinfo{person}{Aylin Caliskan}, \bibinfo{person}{Joanna~J
  Bryson}, {and} \bibinfo{person}{Arvind Narayanan}.}
  \bibinfo{year}{2017}\natexlab{}.
\newblock \showarticletitle{Semantics derived automatically from language
  corpora contain human-like biases}.
\newblock \bibinfo{journal}{\emph{Science}} \bibinfo{volume}{356},
  \bibinfo{number}{6334} (\bibinfo{year}{2017}), \bibinfo{pages}{183--186}.
\newblock


\bibitem[\protect\citeauthoryear{Cer, Yang, Kong, Hua, Limtiaco, John,
  Constant, Guajardo-Cespedes, Yuan, Tar, et~al\mbox{.}}{Cer
  et~al\mbox{.}}{2018}]%
        {cer2018universal}
\bibfield{author}{\bibinfo{person}{Daniel Cer}, \bibinfo{person}{Yinfei Yang},
  \bibinfo{person}{Sheng-yi Kong}, \bibinfo{person}{Nan Hua},
  \bibinfo{person}{Nicole Limtiaco}, \bibinfo{person}{Rhomni~St John},
  \bibinfo{person}{Noah Constant}, \bibinfo{person}{Mario Guajardo-Cespedes},
  \bibinfo{person}{Steve Yuan}, \bibinfo{person}{Chris Tar}, {et~al\mbox{.}}}
  \bibinfo{year}{2018}\natexlab{}.
\newblock \showarticletitle{Universal sentence encoder}.
\newblock \bibinfo{journal}{\emph{arXiv preprint arXiv:1803.11175}}
  (\bibinfo{year}{2018}).
\newblock


\bibitem[\protect\citeauthoryear{Devlin, Chang, Lee, and Toutanova}{Devlin
  et~al\mbox{.}}{2019}]%
        {devlin2018bert}
\bibfield{author}{\bibinfo{person}{Jacob Devlin}, \bibinfo{person}{Ming{-}Wei
  Chang}, \bibinfo{person}{Kenton Lee}, {and} \bibinfo{person}{Kristina
  Toutanova}.} \bibinfo{year}{2019}\natexlab{}.
\newblock \showarticletitle{{BERT:} Pre-training of Deep Bidirectional
  Transformers for Language Understanding}. In
  \bibinfo{booktitle}{\emph{Proceedings of the 2019 Conference of the North
  American Chapter of the Association for Computational Linguistics: Human
  Language Technologies, {NAACL-HLT}}}. \bibinfo{publisher}{Association for
  Computational Linguistics}, \bibinfo{pages}{4171--4186}.
\newblock


\bibitem[\protect\citeauthoryear{Jentzsch, Schramowski, Rothkopf, and
  Kersting}{Jentzsch et~al\mbox{.}}{2019}]%
        {jentzsch2019}
\bibfield{author}{\bibinfo{person}{Sophie Jentzsch}, \bibinfo{person}{Patrick
  Schramowski}, \bibinfo{person}{Constantin Rothkopf}, {and}
  \bibinfo{person}{Kristian Kersting}.} \bibinfo{year}{2019}\natexlab{}.
\newblock \showarticletitle{Semantics Derived Automatically from Language
  Corpora Contain Human-like Moral Choices}. In
  \bibinfo{booktitle}{\emph{Proceedings of the AAAI/ACM Conference on AI,
  Ethics, and Society (AIES)}}. \bibinfo{publisher}{{ACM}},
  \bibinfo{pages}{37--44}.
\newblock


\bibitem[\protect\citeauthoryear{Reimers and Gurevych}{Reimers and
  Gurevych}{2019}]%
        {reimers2019sentence}
\bibfield{author}{\bibinfo{person}{Nils Reimers} {and} \bibinfo{person}{Iryna
  Gurevych}.} \bibinfo{year}{2019}\natexlab{}.
\newblock \showarticletitle{Sentence-BERT: Sentence Embeddings using Siamese
  BERT-Networks}. In \bibinfo{booktitle}{\emph{Proceedings of the 2019
  Conference on Empirical Methods in Natural Language Processing, {EMNLP}}}.
  \bibinfo{publisher}{Association for Computational Linguistics},
  \bibinfo{pages}{3980--3990}.
\newblock


\bibitem[\protect\citeauthoryear{Schramowski, Stammer, Teso, Brugger, Herbert,
  Shao, Luigs, Mahlein, and Kersting}{Schramowski et~al\mbox{.}}{2020a}]%
        {schramowski2020making}
\bibfield{author}{\bibinfo{person}{Patrick Schramowski},
  \bibinfo{person}{Wolfgang Stammer}, \bibinfo{person}{Stefano Teso},
  \bibinfo{person}{Anna Brugger}, \bibinfo{person}{Franziska Herbert},
  \bibinfo{person}{Xiaoting Shao}, \bibinfo{person}{Hans-Georg Luigs},
  \bibinfo{person}{Anne-Katrin Mahlein}, {and} \bibinfo{person}{Kristian
  Kersting}.} \bibinfo{year}{2020}\natexlab{a}.
\newblock \showarticletitle{Making deep neural networks right for the right
  scientific reasons by interacting with their explanations}.
\newblock \bibinfo{journal}{\emph{Nature Machine Intelligence}}
  \bibinfo{volume}{2}, \bibinfo{number}{8} (\bibinfo{year}{2020}),
  \bibinfo{pages}{476--486}.
\newblock


\bibitem[\protect\citeauthoryear{Schramowski, Turan, Jentzsch, Rothkopf, and
  Kersting}{Schramowski et~al\mbox{.}}{2019}]%
        {ArxivBert}
\bibfield{author}{\bibinfo{person}{Patrick Schramowski},
  \bibinfo{person}{Cigdem Turan}, \bibinfo{person}{Sophie Jentzsch},
  \bibinfo{person}{Constantin Rothkopf}, {and} \bibinfo{person}{Kristian
  Kersting}.} \bibinfo{year}{2019}\natexlab{}.
\newblock \showarticletitle{BERT has a Moral Compass: Improvements of ethical
  and moral values of machines}.
\newblock \bibinfo{journal}{\emph{arXiv preprint arXiv:1912.05238}}
  (\bibinfo{year}{2019}).
\newblock


\bibitem[\protect\citeauthoryear{Schramowski, Turan, Jentzsch, Rothkopf, and
  Kersting}{Schramowski et~al\mbox{.}}{2020b}]%
        {Frontiers}
\bibfield{author}{\bibinfo{person}{Patrick Schramowski},
  \bibinfo{person}{Cigdem Turan}, \bibinfo{person}{Sophie Jentzsch},
  \bibinfo{person}{Constantin Rothkopf}, {and} \bibinfo{person}{Kristian
  Kersting}.} \bibinfo{year}{2020}\natexlab{b}.
\newblock \showarticletitle{The Moral Choice Machine}.
\newblock \bibinfo{journal}{\emph{Frontiers in Artificial Intelligence}}
  \bibinfo{volume}{3} (\bibinfo{date}{May} \bibinfo{year}{2020}),
  \bibinfo{pages}{36}.
\newblock


\bibitem[\protect\citeauthoryear{Turan, Neergaard, and Lam}{Turan
  et~al\mbox{.}}{2019}]%
        {turan2019facial}
\bibfield{author}{\bibinfo{person}{Cigdem Turan}, \bibinfo{person}{Karl~David
  Neergaard}, {and} \bibinfo{person}{Kin-Man Lam}.}
  \bibinfo{year}{2019}\natexlab{}.
\newblock \showarticletitle{Facial Expressions of Comprehension (FEC)}.
\newblock \bibinfo{journal}{\emph{IEEE Transactions on Affective Computing}}
  (\bibinfo{year}{2019}).
\newblock


\end{thebibliography}


\end{document}